# The high-pressure synthesis of MgB$_2$.

T. A. Prikhna[1], W. Gawalek[2], A. B. Surzhenko[2], N. V. Sergienko[1], V.E. Moshchil[1], T. Habisreuther[2],

V. S. Melnikov[3], S. N. Dub[1], P. A. Nagorny[1], M. Wendt[2], Ya. M. Savchuk[1], D. Litzkendorf[2],

J. Dellith[2], S. Kracunovska[2], Ch. Schmidt[2]

[1] *Institute for Superhard Materials , 2 Avtozavodskaya St.,Kiev, 04074, Ukraine*
[2] *Institut für Physikalische Hochtechnologie, Winzerlaer Str. 10, Jena, D-07743 Germany*
[3] *Institute of Geochemistry, Mineralogy and Ore-Formation, 34 Palladin Pr., Kiev, 03142, Ukraine*

The high-pressure synthesis is one of the most promising methods to produce MgB$_2$ superconductive (SC) material. The available high-pressure apparatuses have working volume up to 100 cm$^3$ which allows samples up to 60 mm in diameter to be manufactured. By a short-term (1h) high-pressure (2GPa) synthesis from Mg and B we have obtained samples 2.4-2.5 g/cm$^3$ in density, 12.1-15.6 GPa in microhardness (under 4.96 N load), 170-200 kA/cm$^2$ at 10 K and 58-62 kA/cm$^2$ at 20 K in critical current density at a magnetic field of 1T. Samples demonstrated 8.5 T irreversible field at 10 K and 5.5 T at 20 K. The Ta positive influence on SC properties of MgB$_2$ has been observed.

**INTRODUCTION**

MgB$_2$, superconductive properties of which have been recently discovered, can be easily synthesized both under ambient and high pressures. A lot of different methods of MgB$_2$ preparation are known (Table 1). Takano et al.[1] and Jung et al.[2] synthesized samples from B and Mg under high pressures: at 3.5 GPa, 1000 $^o$C, 2h and 3.0 GPa, 950 $^o$C, 2h, respectively. Takano et al.[1] have reported that the highest critical current density (j$_c$) in 1 T field at 20 K was 20 kA/cm$^2$, the irreversible field at 10 K H$_{irr}$ ≈5 T and at 20 K H$_{irr}$ ≈ 3 T. Small single crystals of MgB$_2$ that were extracted from sintered sample exhibited j$_c$≈200 kA/cm$^2$ at 10 K and j$_c$≈100 kA/cm$^2$ at 20K in 1 T field; H$_{irr}$ ≈4 T at 10 K and H$_{irr}$ ≈3.5 T at 20 K.[3] After the proton irradiation the irreversible field of MgB$_2$ single crystals has increased up to 6 T (at 20 K), but the j$_c$ has decreased down to 60 kA/cm$^2$ (in 1T field at 20 K). The samples obtained by a long-term synthesis (700$^o$C - 1000 $^o$C, 14 h) or by intermediate grinding and pressing (600$^o$C, 2 h + 800 $^o$C, 2 h +950$^o$C, 1 h + grinding, pressing + 950$^o$C, 2 h) exhibited Hirr≈ 2.5 - 5.5 T at 20K and the jc of about 40-30 kA/cm$^2$ in 1T field.[5,6] Using high pressure sintering of MgB$_2$ powder, the samples with jc=40 kA/cm$^2$ at 20 K in 1T field have been prepared.[7]

The studies of boron isotope effect[8] have suggested that MgB$_2$ is a conventional phonon mediated BCS superconductor. This has been also supported by band structure calculations.[9] The origin of superconductivity with high T$_c$ can be associated with the light boron atoms whose phonon frequency spectrum plays an important role in enhancing the electron interaction.[5]

Since the upper critical field Hc$_2$ of MgB$_2$ exceeds 10 T at 20 K it is expected that the MgB$_2$ wire can be used for magnetic field generation in cryogenic cooling.[7] Bulk MgB$_2$ is very promising for SC electromotors that could operate under the temperature of liquid hydrogen (~20 K).

We synthesized MgB$_2$ from B and Mg under ambient pressure in an Ar atmosphere and under a pressure of 2 GPa. In the present paper we compare the critical current density versus temperature and magnetic field as well as their mechanical properties (microhardness and fracture toughness) for the samples synthesized under the different conditions. The evidence of the positive effect of the presence of Ta during the high-pressure synthesis of MgB$_2$ on the superconductive properties is discussed.

**EXPERIMENTAL**.

Mg and B have been taken in the stoichiometric ratio of MgB$_2$. Then we mixed and milled the components in a high-speed activator for 1-3 min with steel balls. The obtained powder was compacted into tablets. In our experiments we have used three types of initial boron (two types of amorphous boron of chemical ( 99 %) and commercial (95 %, with some contamination of BN) purity, as well as finely crystalline boron (99 %). All types of boron contained impurity of $H_3BO_3$.

The samples were heated for 4 h in an Ar atmosphere under ambient pressure at 800 $^o$C and for 1 h under 2 GPa at 750 $^o$C, 800 $^o$C, 850 $^o$C, 950 $^o$C or 1000 $^o$C.

The high-pressure conditions have been created inside the high-pressure apparatuses (HPA) of the recessed-anvil and cubic (six punches) types, described elsewhere.[10] The both types of apparatuses are usually used for diamond synthesis. They were slightly modernized for our purposes in order to measure temperature by thermocouples and to prevent the contact of the sample with the graphite heater. The working volume of the biggest cube type HPA is of about 100 cm$^3$ (sample can be up to 60 mm in diameter). In our experiments, the sample was in contact with a compacted powder of hexagonal BN or monoclinic $ZrO_2$ or enveloped in a Ta foil and then placed inside the compacted BN or $ZrO_2$.

The structure of materials was studied using SEM and energy dispersive X-ray analysis. The density was calculated from the results of weighing and measurement of sample dimensions. The $j_c$ was estimated from magnetization hysteresis loops obtained on an Oxford Instruments 3001 vibrating sample magnetometer using Bean's model. Hardness was measured on a Matsuzawa Mod. MXT-70 microhardness tester by a Vickers indenter. The fracture toughness was estimated from the length of the radial cracks emanating from the corners of an indent.[11]

## RESULTS AND DISCUSSION

Table 2 lists the critical current (in zero and 1T fields) and irreversible field at 10 and 20 K values for the samples obtained under different conditions from different types of boron. As Fig.1c and Table 2 show for the amorphous commercially pure boron the synthesis at 800 $^o$C (under 2 GPa) is preferable. But it seems that for samples synthesized from amorphous chemically pure boron and finely crystalline boron somewhat better results have been obtained at the synthesis temperature of 950 $^o$C (see Fig. 1 a,b and Table 2). The increase of synthesis time up to 4 h under 2 GPa pressure led to the amorphisation of the sample structure to the increase of MgO content and resulted in the disappearance of the superconductive properties (compare No 28 with No14, Table 2).

The samples synthesized under the high pressure exhibited higher density (2.4 - 2.5 g/cm$^3$) than those synthesized under ambient (1 atm) pressure of Ar (1.6 g/cm$^3$) and better superconductive properties (see Table 2 and Fig. 2).

We observed the evidence of the Ta positive influence on SC properties of $MgB_2$ (Fig.3, 4). The samples synthesized under the same pressure-temperature-time conditions (2 GPa, 800 $^o$C, 1 h) from the same raw materials but enveloped in a Ta foil showed the higher values of critical current density (in 1 T field) and higher irreversible fields at 20 K than those synthesized in contact with BN or $ZrO_2$ (see Fig. 3, 4 and compare samples 11 with 12 and 19, 22 with 20, 21 in Table 2). In the case of finely crystalline B, such a regularity was not so distinct (compare sample 2 and 3, Table 2), probably, due to the fact that 800 $^o$C is not the optimal synthesis temperature for this type of boron (Fig. 2a). Fig. 5 presents the X-ray patterns of the samples synthesized being enveloped in a Ta foil and placed in BN (No. 20, Table 2) and in contact with BN (No. 5, Table 2). Before the X-ray examination both samples were ground into the powder. In the samples synthesized in a Ta foil, a rather big amount of the foreign phase that can be identified as $Ta_2H$ were present (X-phase in Fig. 5b), while the sample synthesized in contact with BN contained practically pure $MgB_2$. The SEM study shows that Ta is present in the thin surface layer 150 μ in depth only (Fig 6 a, b). This prevents us from the final conclusion that the foreign X-phase is $Ta_2H$. We plan to clear up the situation by conducting the further investigations.

The SC properties of $MgB_2$ can depend upon the amount of the impurity oxygen (MgO) or impurity carbon in the samples structure, upon the grain sizes of the present phases (because the coherence length of $MgB_2$ may be commensurable with grains size), etc. We have certain reasons to believe that by placing a Ta foil between the sample and BN or $ZrO_2$, we can influence the SC properties of $MgB_2$ positively through regulation of the amount of impurities and grain sizes.

Table 3 gives the Vickers microhardness and fracture toughness of $MgB_2$ samples synthesized under high pressure. In many cases we were unable to define the fracture toughness value because there were no cracks from the corners of the indents (under the load of 4.96 N, for example). The absence of cracks under so big a load points to the high level of fracture toughness of the samples. The samples synthesized under ambient pressure were so porous that measurements of their microhardness were impracticable.

SEM study has shown that the structure of high-pressure-synthesized samples included the $MgB_2$ single crystals about 0.1– 10 μm in size (Fig. 6c-f, black grains). Matrix phase was a mixture of $MgB_2$, MgO, $MgH_2$ phases, etc. The grains of the phases that were present in the matrix in many cases were so small (smaller than the diameter of the SEM electron probe) that their quantitative microprobe analysis was impossible. Using energy dispersive X-ray analysis we tried to analyze the difference in the amount of magnesium, boron and impurities of oxygen and carbon in the samples synthesized under the same pressure-temperature conditions, but from the different types of initial boron and in contact with different substances. Unfortunately, we have failed to find a distinct relation between the amount of the elements or impurities and superconductive properties of the samples. It should be mentioned only that the samples with the highest value of critical current density in 1 T field and the highest field of irreversibility at 20 K contained a little bit more of carbon and oxygen impurity than the other samples. Figs. 7 a-e show the structures of the samples arranged in order of increasing critical current density in 1 T field and irreversible fields at 20 K. Analyzing the structure of these $MgB_2$ samples by SEM we have observed (Figs. 7 a-e) that the $j_c$ (in 1T) and $H_{irr}$ at 20 K increase with the size of inclusions that contain oxygen and magnesium as well as with the size of $MgB_2$ single crystals. Samples synthesized for 4 h (Fig. 7 f) have no superconductive properties down to 5 K and in their structures appeared amorphous phase.

In the same time the $j_c$ in zero magnetic field is much higher in the case of finer grains and homogeneous structure. X-ray study shows that the samples synthesized in contact with BN from the finely crystalline boron and amorphous boron of the chemical purity in the 750-950 °C range contained mainly $MgB_2$ phase. Such samples have the highest values of $j_c$ at 10 K in zero field: 565-213 kA/cm$^2$, but their irreversible field does not exceed 4.3-6.4 T. At 20 K their SC properties decrease very abruptly (see Table 2).

The samples synthesized from boron of commercial purity in contact with BN or $ZrO_2$ contained a higher amount of MgO and Mg than the samples synthesized from finely crystalline and amorphous boron. These samples had jc at 10 K in zero field of about 125-159 kA/cm$^2$, but the irreversible field up to 8.5 T. And at 20 K their jc and fields of irreversibility are much higher (up to 30 kA/cm$^2$ in 1 T and Hirr=5.3). The best SC properties at 20 K have been shown by samples synthesized in contact with Ta from boron of commercial purity. The critical current density and irreversible field were approximately the same as for the irradiated $MgB_2$ single crystals [4]. The X-ray examination (Fig. 5 b) has shown a higher amount of MgO and Mg impurities, the presence of $MgH_2$ and X-phase (Ta$_2$H?).

Homogeneous high-density $MgB_2$ that has been high-pressure synthesized from finely crystalline boron and amorphous chemically pure boron can be used as target for film deposition. The high SC characteristics (high critical current density in magnetic field and high irreversibility fields) of the composition based on $MgB_2$ (synthesized from magnesium and boron of commercial purity) make this material promising for the using in SC electrical machines. It should be mentioned that for cryogenic applications it is important to trap a high magnetic energy in the material or to get a high current load that is the product of the critical current density and the diameter of the current loop in the sample. So, the possibility to manufacture large samples using high-pressure technique makes this method suitable for the commercial production of $MgB_2$ for cryogenic purposes.

**CONCLUSIONS**

1. High-pressure synthesis of highly dense $MgB_2$ seems to be a very promising method of manufacturing this material. Moreover, because large working volume high-pressure apparatuses can be used.
2. By short-term synthesis (for about 1h) under 2 GPa, from Mg and B the samples with 2.4-2.5 g/cm$^3$ density, 12.1-15.6 GPa Vickers microhardness (under 4.96 N load) and high fracture toughness have been obtained. The critical current density of the samples synthesized from Mg and B of commercial purity in 1 T magnetic field was 170-200 kA/cm2 (at 10 K) and 58-62 (at 20 K), the irreversible field was 8.5 and 5.5 T at 10 and 20 K, respectively.
3. The positive influence of the presence of Ta during high-pressure synthesis of $MgB_2$ has been observed on the SC properties of $MgB_2$ at 20 K in 1T magnetic field. The peculiarities of $MgB_2$ structure and properties variation versus conditions of synthesis have been investigated. For deeper understanding of the mechanism of the material structure formation and correlation between structure and properties, a more detailed study should be conducted.

4. The samples of homogeneous $MgB_2$ demonstrate high critical current density in zero magnetic field, while the critical current density of inhomogeneous $MgB_2$-based samples was higher in 1 T field at 20 K, as well as their fields of irreversibility, than that of homogeneous $MgB_2$ samples of a low impurity content. The similar behavior is usually observed on other superconductors.

TABLE 1 Critical current density, $j_c$, (in zero and 1T magnetic field) and irreversible field, $H_{irr}$, versus temperature for $MgB_2$ samples synthesized under different conditions.

| Author, [Ref. No]; P, T, τ, conditions of synthesis or sintering | at 10 K | | | at 20 K | | |
|---|---|---|---|---|---|---|
| | $j_c$, kA/cm² | | $H_{irr}$, T | $j_c$, kA/cm² | | $H_{irr}$, T |
| | 0 T | 1 T | | 0 T | 1 T | |
| Y. Takano et al. [1], 3.5 GPa, 1000 °C, 2 h | - | - | 5 | - | 20 | 3 |
| Y. Bogoslavsky et al [3], Singlecrystalline $MgB_2$ | 500 | 200 | 4 | 400 | 100 | 3.5 |
| Y. Bogoslavsky et al. [4] Singlecrystalline irradiated $MgB_2$ | - | - | 9.5 | 100 | 60 | 6 |
| G. Amish et al. [5], 600°C, 2 h + 800 °C, 2 h +950°C, 1 h + grinding, pressing + 950°C, 2 h | 900 | 150 | 4 | 480 | 40 | 2.5 |
| S.X. Dou et al., [6] 700-1000 °C, 1-14 h. | 65 | 60 | 7.5 | 90 | 30 | 5.5 |
| K. Togano et al., [7] High pressure sintered $MgB_2$ | - | - | - | - | 40 | - |

TABLE 2 Critical current density (in zero and 1T field) and irreversible field versus temperature for the samples synthesized under different high-pressure conditions from boron of different types.

| No | Regime: P, T, τ | Type of Boron* | In contact with | 10 K Jc, kA/cm² 0 T | 10 K Jc, kA/cm² 1 T | 10 K $H_{irr}$ | 20 K Jc, kA/cm² 0 T | 20 K Jc, kA/cm² 1 T | 20 K $H_{irr}$ |
|---|---|---|---|---|---|---|---|---|---|
| 1 | 1 atm, 800 °C, 4 h | Finely cr. | Ar | 93 | 22 | 4.6 | 26 | 0.6 | 1.96 |
| 2 | 2 GPa, 800 °C, 1 h | Finely cr. | BN | 213 | 113 | 5.3 | 39 | 3 | 2.2 |
| 3 | 2 GPa, 800 °C, 1 h | Finely cr. | Ta+BN | 207 | 119 | 5 | 43 | 2.5 | 3.9 |
| 4 | 2 GPa, 800 °C, 1 h | Finely cr. | $ZrO_2$ | 234 | 120 | 4.3 | 43 | 2 | 2.0 |
| 5 | 2 GPa, 750 °C, 1 h | Finely cr. | BN | 251 | 133 | 5.6 | 48 | 5.5 | 2.5 |
| 6 | 2 GPa, 850 °C, 1 h | Finely cr. | BN | 219 | 120 | 4.6 | 40 | 17 | 1.6 |
| 7 | 2 GPa, 900 °C, 1 h | Finely cr. | BN | 565 | 280 | 5.4 | 160 | 19 | 2.56 |
| 8 | 2 GPa, 950 °C, 1 h | Finely cr. | BN | 337 | 166 | 6.4 | 138 | 30 | 3.78 |
| 9 | 2 GPa, 1000 °C, 1 h | Finely cr. | BN | No SC properties | | | | | |
| 10 | 1 atm, 800 °C, 4 h | Am. | Ar | 27 | 5.5 | 3.36 | 2.5 | 0.06 | 2.86 |
| 11 | 2 GPa, 800 °C, 1 h | Am | BN | 279 | 118 | 4.5 | 35 | 1.4 | 1.7 |
| 12 | 2 GPa, 800 °C, 1 h | Am | Ta+BN | 364 | 200 | 4.5 | 80 | 47 | 4.3 |
| 13 | 2 GPa, 750 °C, 1 h | Am | BN | 187 | 75 | 4.6 | 20 | 1.3 | 2.0 |
| 14 | 2 GPa, 850 °C, 1 h | Am | BN | 241 | 85 | 3.9 | 22 | 0.308 | 1.3 |
| 15 | 2 GPa, 900 °C, 1 h | Am | BN | 87 | 26 | 2.9 | 3 | 0.003 | 0.8 |
| 16 | 2 GPa, 950 °C, 1 h | Am | BN | 257 | 12 | 4.5 | 92 | 11 | 2.6 |
| 17 | 2 GPa, 1000 °C, 1 h | Am | BN | No SC properties | | | | | |
| 18 | 1 atm, 800 °C, 4 h | Am.(com) | Ar | 133 | 44 | 4.52 | 69 | 8 | 2.8 |
| 19 | 2 GPa, 800 °C, 1 h | Am.(com) | BN | 125 | 79 | 8.5 | 69 | 30 | 5.3 |
| 20 | 2 GPa, 800 °C, 1 h | Am.(com) | Ta+BN | 210 | 162 | 7.9 | 115 | 62 | 5.2 |
| 21 | 2 GPa, 800 °C, 1 h | Am.(com) | Ta+$ZrO_2$ | 231 | 170 | 7.9 | 131 | 58 | 5.5 |
| 22 | 2 GPa, 800 °C, 1 h | Am.(com) | $ZrO_2$ | 159 | 54 | 4.6 | 65 | 8 | 2.6 |
| 23 | 2 GPa, 750 °C, 1 h | Am.(com) | BN | 17 | 7.3 | 8 | 7 | 1.9 | 5.0 |
| 24 | 2 GPa, 850 °C, 1 h | Am.(com) | BN | 43 | 15 | 3.94 | 49 | 12 | 3.4 |
| 25 | 2 GPa, 900 °C, 1 h | Am.(com) | BN | 33 | 6.2 | 4.8 | 18 | 2 | 3 |
| 26 | 2 GPa, 950 °C, 1 h | Am.(com) | BN | 232 | 68 | 4.2 | 120 | 15 | 2.6 |
| 27 | 2 GPa, 1000 °C, 1 h | Am.(com) | BN | 23.4 | 0.1 | 1.4 | - | - | - |
| 28 | 2 GPa, 850 °C, 4 h | Am | BN | No SC properties | | | | | |

* -Finely cr. –finely crystalline boron, Am. – amorphous chemically pure boron, Am. (com..) – amorphous boron of commercial purity.

TABLE 3. Vickers microhardness and fracture toughness of the synthesized samples

| No | Regime: P, T, τ | Type of boron | In contact with | $H_V$, GPa at 4.96 N | $K_{1C}$, MN×m$^{-3/2}$, at 4.96 N |
|---|---|---|---|---|---|
| 1 | 2 GPa, 800 °C, 1 h | Finely cr. | BN | 15.03±0.68 | 3.64±0.52 |
| 2 | 2 GPa, 800 °C, 1 h | Finely cr. | Ta+BN | 14.58± 0.53 | No cracks |
| 3 | 2 GPa, 750 °C, 1 h | Finely cr. | BN | 13.99±0.78 | 3.87±0.79 |
| 4 | 2 GPa, 900 °C, 1 h | Finely cr. | BN | 14.10±0.41 | No cracks |
| 5 | 2 GPa, 950 °C, 1 h | Finely cr. | BN | 12.12±0.87 | No cracks |
| 6 | 2 GPa, 800 °C, 1 h | Am | Ta+BN | 12.79±1.7 | No cracks |
| 7 | 2 GPa, 800 °C, 1 h | Am | $ZrO_2$ | 11.25±0.27 | No cracks |
| 8 | 2 GPa, 750 °C, 1 h | Am | BN | 11.35±1.13 | 3.00±0.32 |
| 9 | 2 GPa, 850 °C, 1 h | Am | BN | 12.98±0.45 | 2.72±0.24 |
| 10 | 2 GPa, 850 °C, 4 h | Am | BN | 13.1±1.4 | No cracks |
| 11 | 2 GPa, 900 °C, 1 h | Am | BN | 11.94±0.93 | No cracks |
| 12 | 2 GPa, 950 °C, 1 h | Am | BN | 13.97±0.29 | 4.24±0.14 |
| 13 | 2 GPa, 900 °C, 1 h | Am.(com) | BN | 12.79±1.14 | No cracks |
| 14 | 2 GPa, 950 °C, 1 h | Am.(com) | BN | 7.94±0.84 | No cracks |

**Figure Captions**

FIG.1 Critical current density, $j_c$, in zero and 1 T magnetic field and irreversible field, $H_{irr}$, at 10 and 20 K versus the synthesis temperature, $T_s$, of the samples synthesized under 2 GPa for 1 h in contact with BN from Mg and different type of B: (a) – finely crystalline, (b) – amorphous chemically pure,  (c) – amorphous boron of commercial purity.

FIG.2 Critical current density (at 10 K) vs. magnetic field and irreversible field vs. temperature for the samples synthesized from amorphous commercially pure boron under ambient pressure (in Ar) and under a high pressure ( in contact with a Ta foil, placed in $ZrO_2$ ), Nos. 18 and 21, Table 2, respectively.

FIG. 3. Critical current density, $j_c$, at different temperatures vs. magnetic field, $\mu_0H$, for the samples synthesized from Mg and amorphous commercially pure B under the same pressure-temperature-time conditions (a) enveloped in Ta in contact with $ZrO_2$, (sample 21, Table 2) (b) in contact with BN (sample 19, Table 2). Inside each graph the dependence of irreversible field, $\mu_0H_{irr}$, on temperature, T, is shown.

FIG. 4. Critical current density, $j_c$, at 20 K vs. magnetic field, $\mu_0H$, for the samples synthesized from Mg and amorphous commercially pure B at 800 °C in Ar atmosphere for 4 h (sample 18, Table 2) and under 2 GPa for 1h enveloped in Ta in contact with BN (sample 19, Table 2) and $ZrO_2$, (sample 21, Table 2) and in contact with $ZrO_2$ (sample 22, Table 2). The inset shows a part of the $j_c$ dependence on. magnetic field ( from 0 up to 1 T) in a linear $j_c$ scale.

FIG. 5  X-ray patterns of the samples synthesized under 2 GPa for 1h (a) from finely crystalline boron at 750 °C in contact with BN (No. 5 , Table 2) and (b) from amorphous commercially pure boron enveloped in Ta in contact with BN at 800 °C (No. 20, Table 2).

FIG. 6 Structure of the samples (obtained by SEM) synthesized under high pressure
(a, b) from Mg and amorphous commercially pure B in contact with Ta, placed in $ZrO_2$ at 2 GPa , 800 °C, for 1h (No. 21, Table 2), (c, d) from Mg and amorphous chemically pure B in contact with BN at 2 GPa, 850 °C, for 1h (No. 14, Table 2): (c)- SEI (secondary electron image), (d) – COMPO (back scattering electron image, composition), (e, f) from Mg and finely crystalline B in contact with BN at 2 GPa, 800 °C, for 1h (No 2, Table 2).

FIG. 7 The structure (by SEM) of the samples synthesized under  2 GPa, 800 °C, for 1h (a-e) from different initial boron in contact with different substances arranged in order of increasing critical current density, $j_c$, in 1 T field and irreversible field, $H_{irr}$, at 20 K  and of the sample synthesized at 2 GPa, 850 °C, for 4 h (f), from chemically pure amorphous B in contact with BN that have no superconductive properties down to 5 K.

| Figure | Initial boron | Synthesized in contact with | T=20K | | Nos acc. to Table 2 |
|---|---|---|---|---|---|
| | | | $j_c$, kA/cm$^2$, in 1 T field | $H_{irr}$, T | |
| 7a | Am. | BN | 1,4 | 1.8 | 11 |
| 7b | Finely. cr. | BN | 3 | 2.2 | 2 |
| 7c | Am.(com) | $ZrO_2$ | 8 | 2.6 | 22 |
| 7d | Am.(com) | BN | 30 | 5.3 | 19 |
| 7e | Am.(com) | Ta+ $ZrO_2$ | 58 | 5.5 | 21 |
| 7f | Am. | BN | No SC properties | | 28 |

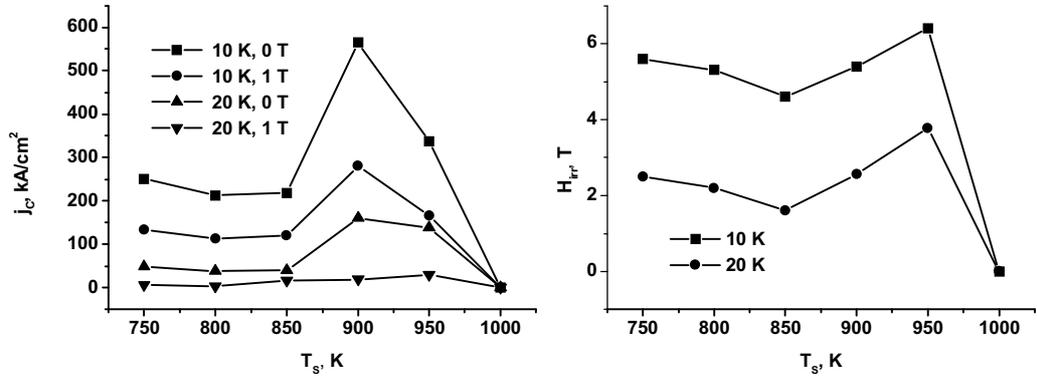

a)

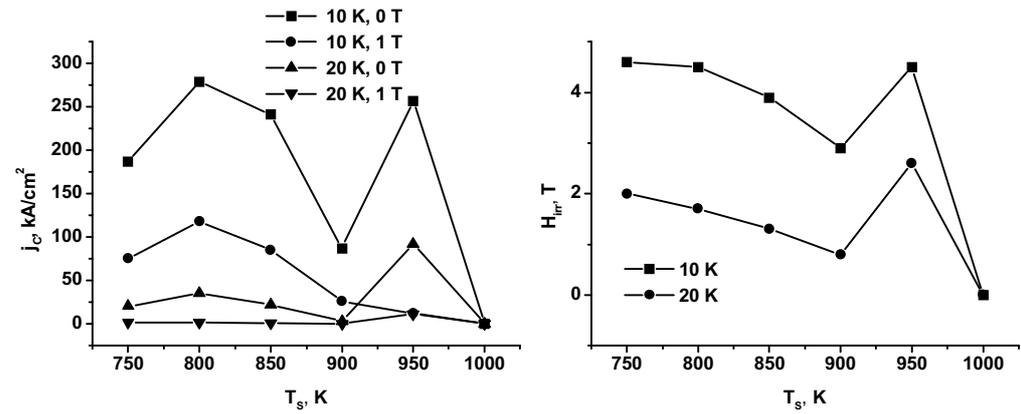

b)

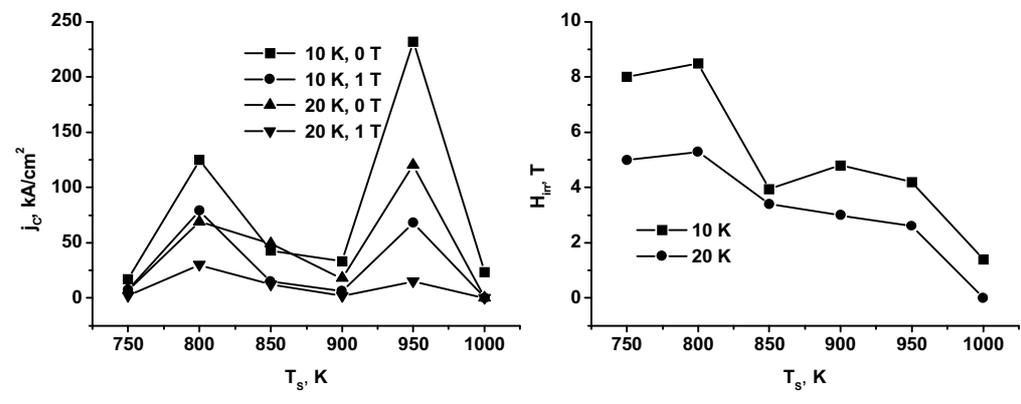

c)

Fig.1

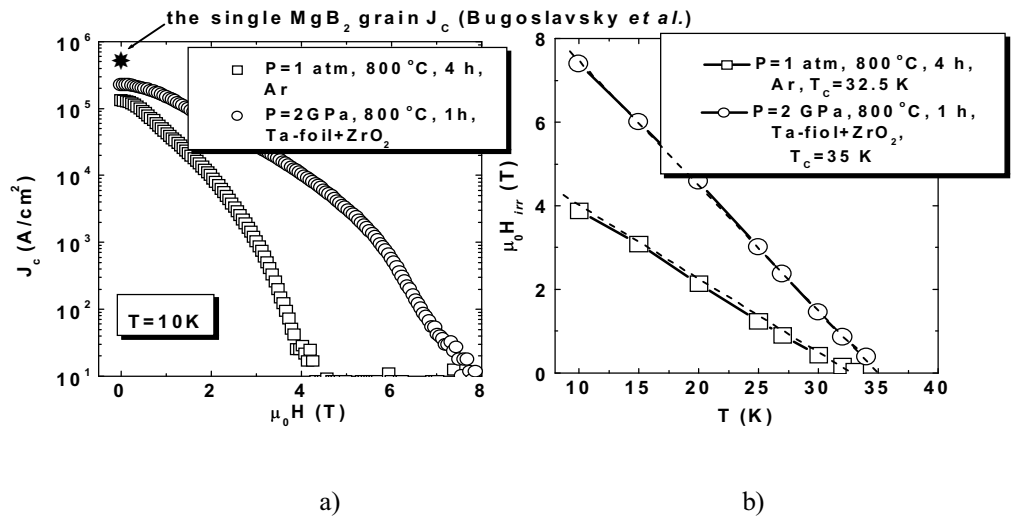

Fig.2

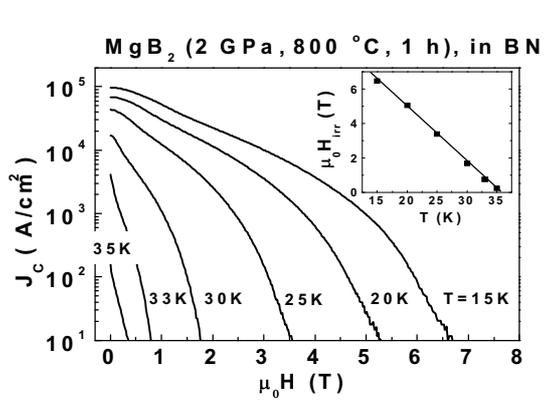 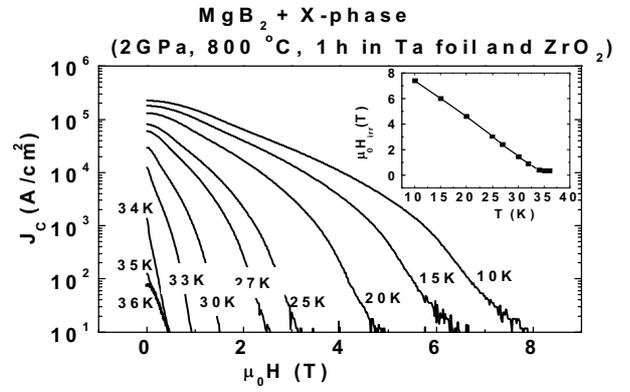

a)                                                            b)

Fig.3

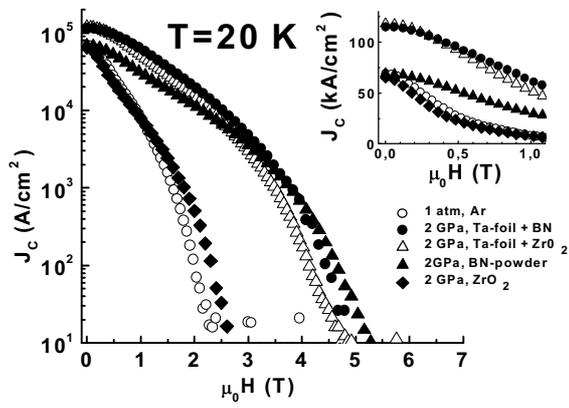

Fig.4

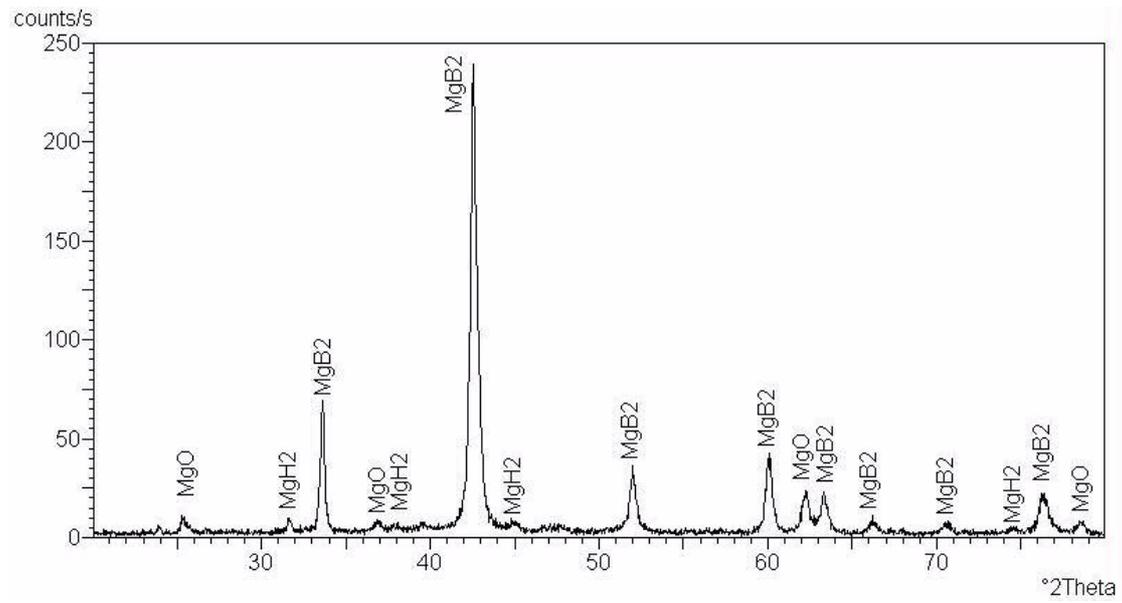

a)

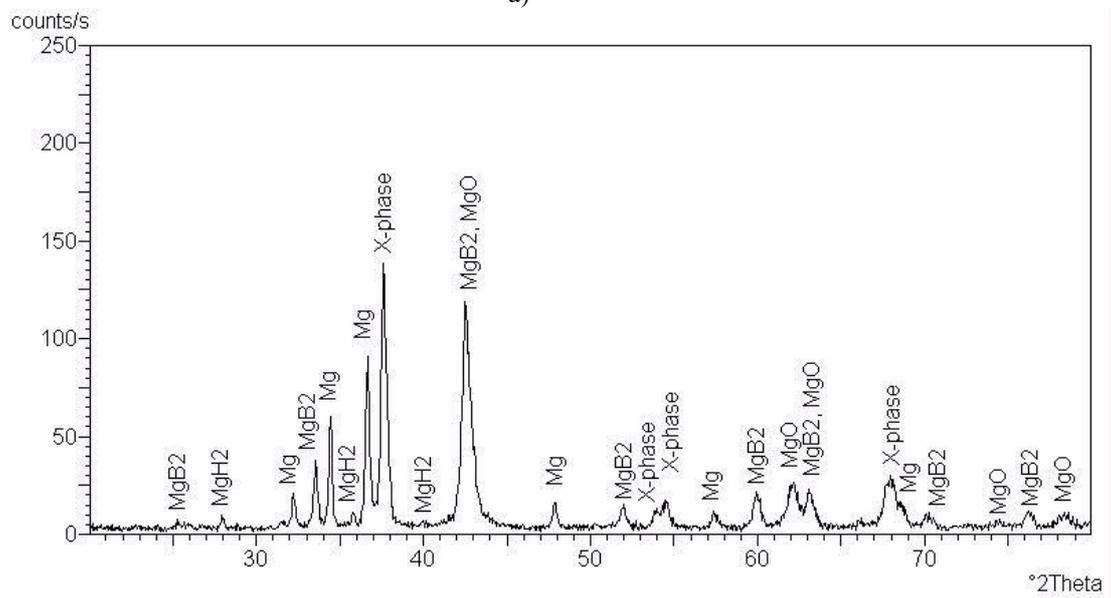

b)

Fig.5

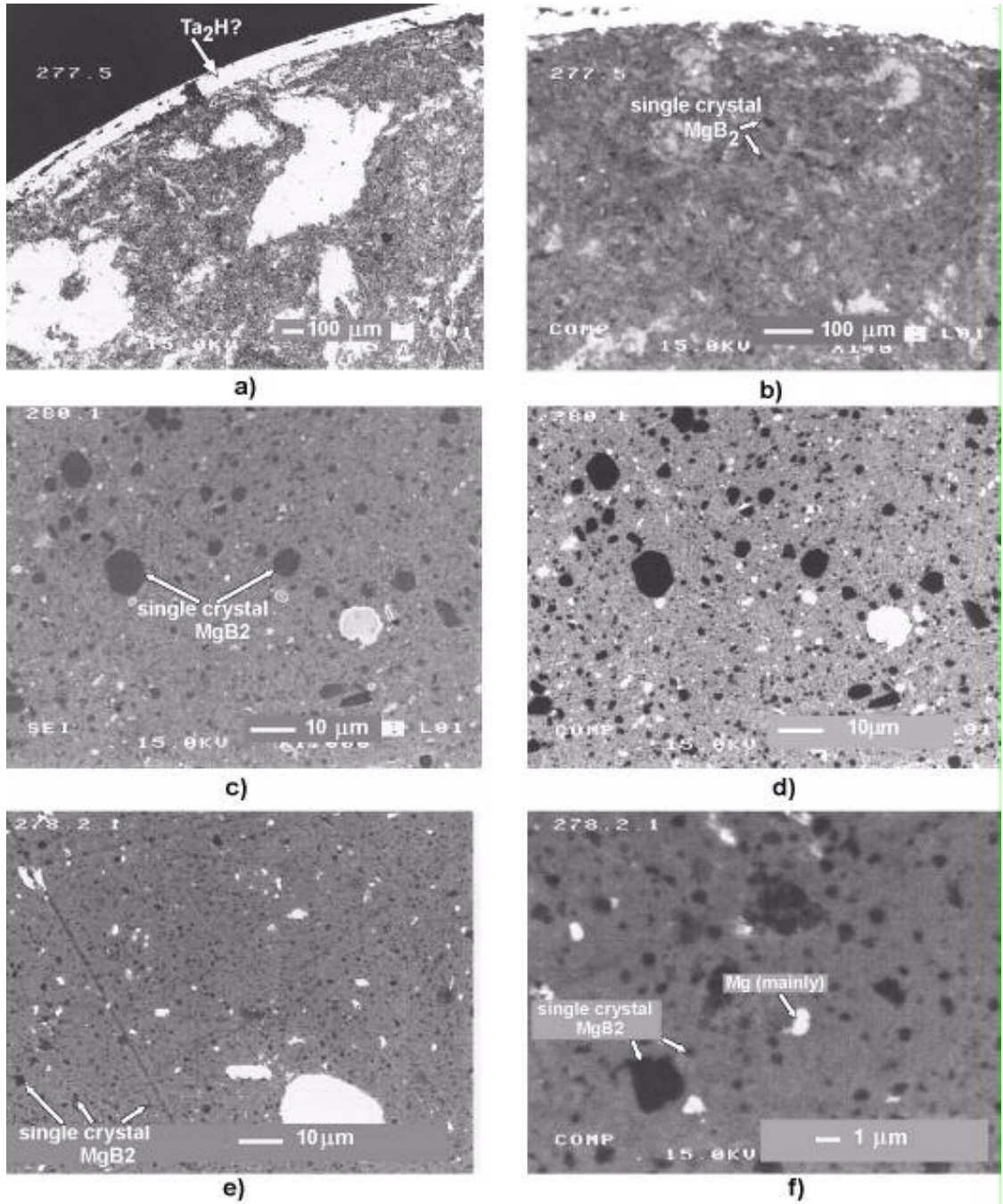

Fig.6

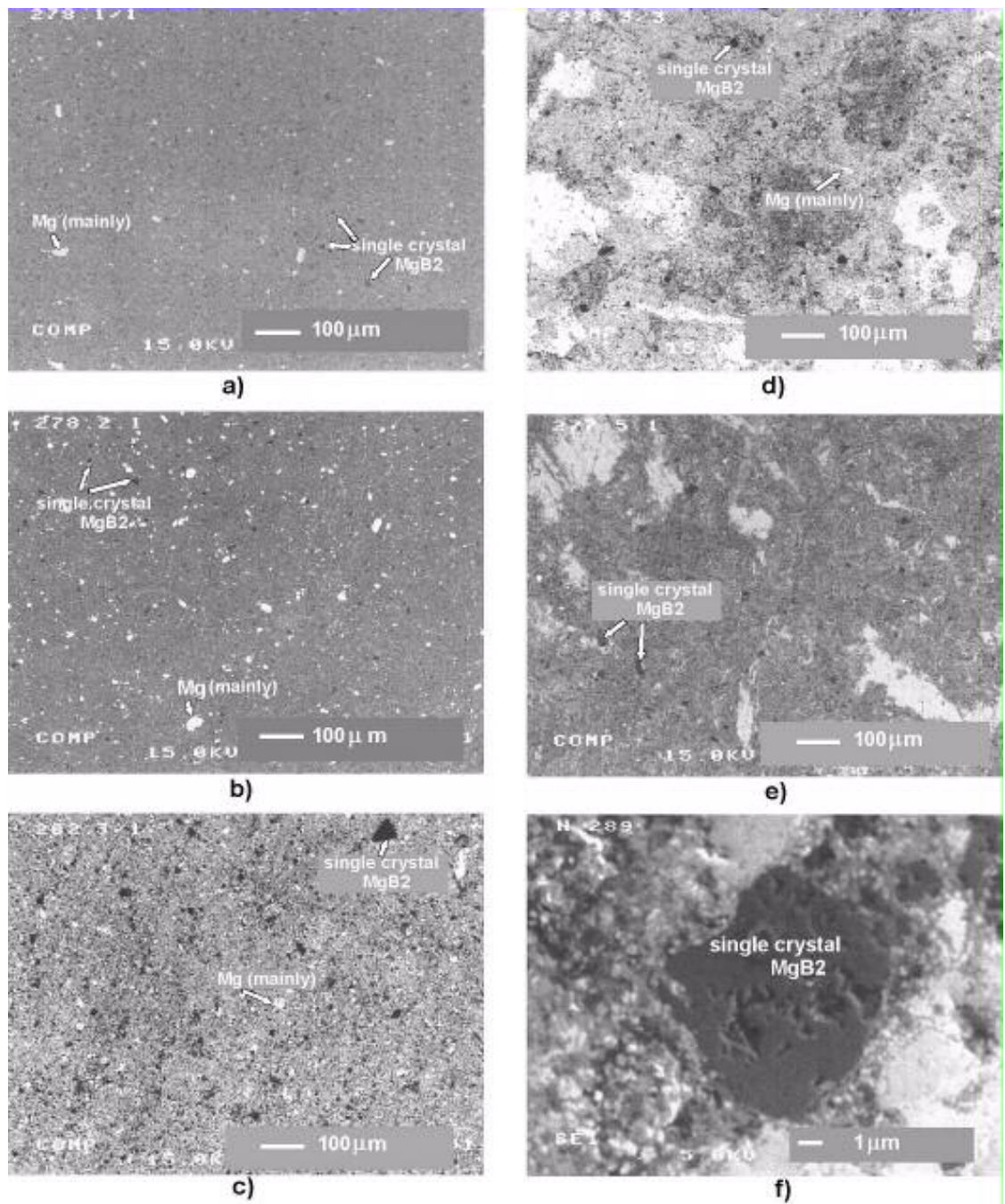

Fig.7